\documentclass{mn2e}
\input{epsf}

\title[Mass distribution in nearby Abell clusters]{Mass distribution in nearby Abell clusters}

\author[E. L. {\L}okas et al.]
    {E. L. {\L}okas,$^{1}$\thanks{E-mail: lokas@camk.edu.pl}
    R. Wojtak,$^{2}$
    S. Gottl\"ober,$^{3}$
    G. A. Mamon$^{4,5}$ and
    F. Prada$^{6}$
    \\
    \\
    $^1$Nicolaus Copernicus Astronomical Center, Bartycka 18,
    00-716 Warsaw, Poland \\
    $^2$Astronomical Observatory, Jagiellonian University, Orla 171, 30-244 Cracow,
    Poland \\
    $^3$Astrophysikalisches Institut Potsdam, An der Sternwarte 16, 14482 Potsdam, Germany\\
    $^4$Institut d'Astrophysique de Paris (UMR 7095: CNRS and Universit\'e Pierre \& Marie Curie),
    98 bis Bd Arago,
    F-75014 Paris, France\\
    $^5$GEPI (UMR 8111: CNRS and Universit\'e Denis Diderot), Observatoire de Paris,
    F-92195 Meudon, France\\
    $^6$Instituto de Astrof{\'\i}sica de Andalucia (CSIC),
    Apartado Correos 3005, E-18080 Granada, Spain}
\begin{document}

\maketitle

\begin{abstract}
We study the mass distribution in six nearby ($z<0.06$) relaxed Abell clusters of galaxies
A0262, A0496, A1060, A2199, A3158 and A3558. Given the dominance of
dark matter in galaxy clusters we approximate their total density distribution
by the NFW formula characterized by virial mass and concentration. We also assume
that the anisotropy of galactic orbits is reasonably well described by a constant and
that galaxy distribution traces that of the total density.
Using the velocity and position data for 120-420 galaxies per cluster
we calculate, after removal of interlopers, the profiles of the lowest-order
even velocity moments, dispersion and
kurtosis. We then reproduce the velocity moments by jointly fitting the moments
to the solutions of the Jeans equations.
Including the kurtosis in the analysis allows us to
break the degeneracy between the mass distribution and anisotropy and
constrain the anisotropy as well as the virial mass and concentration.
The method is tested in detail on mock data extracted from $N$-body simulations of
dark matter haloes.
We find that the best-fitting galactic orbits are remarkably close to isotropic in most
clusters. Using the fitted pairs of mass and
concentration parameters for the six clusters we conclude that the trend of decreasing
concentration for higher masses found
in cosmological $N$-body simulations is consistent with the data.
By scaling the individual cluster data by mass we combine them to create a composite cluster
with 1465 galaxies and perform a similar analysis on such sample. The estimated concentration
parameter then lies in the range $1.5 < c < 14$ and the anisotropy parameter
in the range $-1.1 < \beta < 0.5$ at the 95 percent confidence level.
\end{abstract}


\section{Introduction}

Studies of galaxy kinematics in clusters remain a major tool in determining the mass
distribution in these objects, complemented by methods based on the analysis
of the hot X-ray gas and gravitational lensing. Due to a limited number of
measured galaxy redshifts per
cluster such analyses have been usually performed on composite clusters by combining data from
many objects (e.g. Carlberg et al. 1997; van der Marel et al. 2000; Biviano \& Girardi 2003;
Mahdavi \& Geller 2004; Katgert, Biviano \& Mazure 2004; Biviano \& Katgert 2004; Goto 2005;
Biviano \& Salucci 2005).
The normalizations needed in stacking the clusters together reduce however the number of
parameters that
can be estimated from the analysis. Besides, such studies are usually restricted to the analysis
of velocity dispersion profile with the simplifying assumption of isotropic galactic orbits.

\begin{table*}
\begin{center}
\caption{Observational parameters of the clusters. }
\begin{tabular}{cccccccccc}
assigned & cluster & RA     & Dec     & redshift & number      & velocity dispersion  & kurtosis \\
number	 & name    & (J2000) & (J2000) & $z$      & of galaxies & $\sigma_{\rm los}$ [km s$^{-1}$] & $\kappa_{\rm los}$ \\
\hline
1 & A0262  & $01^{\rm h}52^{\rm m}50.4^{\rm s}$ & $+36^\circ 08'46''$ & 0.0163 & 120          &  527 $\pm$ 34 &  2.63 $\pm$ 0.44  \\
2 & A0496  & $04^{\rm h}33^{\rm m}37.1^{\rm s}$ & $-13^\circ 14'46''$ & 0.0329 & 270          &  719 $\pm$ 31 &  3.36 $\pm$ 0.30  \\
3 & A1060  & $10^{\rm h}36^{\rm m}51.3^{\rm s}$ & $-27^\circ 31'35''$ & 0.0126 & 330          &  696 $\pm$ 27 &  2.68 $\pm$ 0.27  \\
4 & A2199  & $16^{\rm h}28^{\rm m}37.0^{\rm s}$ & $+39^\circ 31'28''$ & 0.0302 & 180          &  795 $\pm$ 42 &  2.49 $\pm$ 0.36  \\
5 & A3158  & $03^{\rm h}42^{\rm m}39.6^{\rm s}$ & $-53^\circ 37'50''$ & 0.0597 & 145          &  970 $\pm$ 57 &  2.58 $\pm$ 0.40  \\
6 & A3558  & $13^{\rm h}27^{\rm m}54.8^{\rm s}$ & $-31^\circ 29'32''$ & 0.0480 & 420          &  948 $\pm$ 33 &  2.70 $\pm$ 0.24  \\
\hline
\label{properties}
\end{tabular}
\end{center}
\end{table*}

In this study we attempt a kinematical analysis of six individual clusters which we supplement
in the end by a similar procedure performed on a composite cluster created from the galaxies
belonging to the six clusters. Our method relies on an extension of the usual Jeans formalism
beyond the lowest-order velocity moment and including also the kurtosis of the line-of-sight
velocity distribution (\L okas 2002; \L okas \& Mamon 2003). The formalism
has been successfully applied to study the dark matter distribution in the Coma cluster
of galaxies by \L okas \& Mamon (2003). It has been shown that, for a restricted class of
dark matter distributions motivated by the results of cosmological $N$-body simulations,
the joint analysis of velocity
dispersion and kurtosis allows us to break the usual degeneracy between the mass
distribution and velocity anisotropy and constrain the parameters of the dark matter
profile. Recently we have tested the reliability of this approach against a series
of $N$-body simulations (Sanchis, \L okas \& Mamon 2004) and also applied it
to constrain the dark matter distribution in the Draco dwarf spheroidal
galaxy (\L okas, Mamon \& Prada 2005a).

Here we further test the method on a different set of cosmological $N$-body simulations
by studying in detail the errors in the estimated parameters
following from the sampling errors of velocity moments. We also
introduce and test a new procedure of interloper removal which we then apply to the clusters.
With a number of available galaxy redshifts per cluster much smaller than for Coma we had however
to introduce a number of simplifications in our modelling compared to \L okas \& Mamon (2003).
We model the total mass distribution instead of only the dark matter component, we assume that
galaxies trace the total mass distribution and we use the velocity data of all galaxies, not
only ellipticals. On the other hand, we believe that the carefully selected clusters studied here
are much more relaxed compared to Coma and therefore we avoid any uncertainties due to
departures from dynamical equilibrium.

The paper is organized as follows. In Section~2 we briefly describe our data set. In Section~3
we summarize our method of data modelling including the removal of interlopers and
fitting of velocity moments and test it on a sample of dark matter haloes extracted from
a cosmological $N$-body simulation. The results for the six galaxy clusters and the composite
cluster are presented in Section~4. The discussion follows in Section~5.

\section{The data}

We have searched the NASA/IPAC Extragalactic Database (NED) for nearby ($z < 0.1$)
well-studied galaxy clusters with
at least 120 galaxies within projected distance of about 2 Mpc from the cluster centre and
with $c z$ velocities differing from the cluster mean by less than $\pm 4000$ km s$^{-1}$.
Among a few tens of clusters selected in this way we have chosen those which are likely to
be relaxed judging by the regularity of their X-ray surface brightness
maps obtained with Einstein (Jones
\& Forman 1999) and ASCA (Horner et al. 2000) satellites. Next we have looked at the
regularity of the diagrams showing the line-of-sight velocities of galaxies
as a function of projected distance from the cluster centre. We have rejected those with
irregular diagrams which may indicate merging or presence of neighbouring structures
({\L}okas et al. 2005b) for which our method of interloper removal does not work.

Table~\ref{properties} lists the clusters we chose for the analysis together with their
positions and redshifts as given by NED. We also give the number of galaxies per cluster
used for the calculation of the velocity moments (the numbers are for the final samples,
after removal of interlopers). The last two columns give the values of
the line-of-sight velocity moments, dispersion and kurtosis calculated for all galaxies.
The method of estimating these values is discussed in the next section. The list
includes some very well known clusters. A0496 was studied in detail by
Durret et al. (2000) who called it a prototype of a relaxed cluster. Indeed the cluster
seems to have the most regular X-ray luminosity distribution of those in our sample
and a single central galaxy whose position coincides with the centre of the gas
distribution. The cluster A1060 (Hydra I, Fitchett \& Merritt 1988),
although quite similar to A0496 in many aspects,
has two central galaxies and somewhat less uniform X-ray distribution so it may have
just reached equilibrium after a major merger. None of the clusters, however, is
completely free of substructure (see e.g. Escalera et al. 1994) or neighbours (even A0496).
A3558 (Shapley 8) is the richest member of the Shapley
supercluster (Dantas et al. 1997; Bardelli et al. 1998)
including a much smaller neighbour A3556 and a bigger, more distant cluster A3562. For this cluster
we have restricted the analysis to distances where members of  A3556 and A3562 are not likely
to contaminate the sample. A2199 seems to be quite a relaxed cluster but has a less massive
neighbour A2197 and other groups (Rines et al. 2001, 2002) which made us restrict the
analysis to distances much smaller than the estimated virial radius. Also A0262, a member of
the Perseus-Pisces supercluster, can be affected by neighbouring structures.

It is generally believed that elliptical galaxies in clusters comprise a virialized, more relaxed
subsample compared to (late) spirals which might be infalling into the cluster for the first time.
It could therefore be desirable to restrict the analysis to ellipticals as was done in the case of
Coma cluster by {\L}okas \& Mamon (2003). However, for the present sample of clusters the
morphological information is available only for a small fraction of galaxies making such an analysis
impossible. Given a larger number of galaxies per cluster one could also attempt to measure their
surface density distribution reliably and use it as an input in the kinematical analysis.
With scarce samples presently available we are forced to
assume that the number density of galaxies follows the total density distribution.

For the kinematical analysis we have chosen as the centres of the clusters their central cD
galaxies which coincide with the centre of the X-ray surface brightness distribution. In the case
of the presence of two central galaxies (in A1060 and A3158) we have chosen as a centre the position
of the one which is closer to the centre of the X-ray surface brightness distribution. The
galaxy velocities have been transformed to the reference frame of the cluster and in order to
calculate the distances within clusters we have transformed the cluster velocities to the
reference frame of the cosmic microwave background. The concordance cosmological model
($\Lambda$CDM) with parameters $\Omega_M=0.3$, $\Omega_{\Lambda}=0.7$ and $h=0.7$ is assumed
throughout the paper.

\section{The method}

\subsection{Overview of the method}

In this section we summarize our method of determining the mass distribution from
velocity moments, as developed in \L okas (2002) and \L okas \& Mamon (2003),
and test it against mock data obtained from $N$-body simulations. The method relies
on fitting the solutions of the Jeans equations for the second and fourth velocity
moments to the profiles of the moments determined from the data.
For the velocity dispersion projected along the line of sight, $\sigma_{\rm los}$, the
Jeans formalism gives (Binney \& Mamon 1982)
\begin{equation}         \label{proj1}
        \sigma_{\rm los}^2 (R) = \frac{2}{I(R)} \int_{R}^{\infty}
	\frac{\nu \sigma_r^2 r}{\sqrt{r^2 - R^2}} \left( 1-\beta \frac{R^2}{r^2} \right)
	{\rm d} r  \ ,
\end{equation}
where $\nu(r)$ and $I(R)$ are the 3D and the surface distribution of the tracer as a function
of a true ($r$) and projected ($R$) distance from the object centre respectively. The
parameter
\begin{equation}	\label{beta}
	\beta=1-\frac{\sigma_\theta^2(r)}{\sigma_r^2(r)}
\end{equation}
describes a relation between the angular $\sigma_\theta$ and radial
$\sigma_r$ velocity dispersions and characterizes the anisotropy of the tracer orbits.
We will assume it here to be constant with radius and consider $-\infty < \beta \le 1$
which covers all interesting possibilities from radial orbits
($\beta=1$) to isotropy ($\beta=0$) and circular orbits
($\beta \rightarrow - \infty$). For constant $\beta$ the radial velocity dispersion,
$\sigma_r(r)$ in equation (\ref{proj1}) is
\begin{equation}           \label{sol1}
	\nu \sigma_r^2 (r) = r^{-2 \beta}
	\int_r^\infty r^{2 \beta} \nu \frac{{\rm d} \Phi}{{\rm d} r} \ {\rm d}r
\end{equation}
where $\Phi$ is the gravitational potential.

For the fourth projected velocity moment we have
\begin{equation}             \label{proj2}
	\overline{v_{\rm los}^4} (R) = \frac{2}{I(R)} \int_{R}^{\infty}
	\frac{\nu \,  \overline{v_r^4} \,r}{\sqrt{r^2 - R^2}} \ g(r, R, \beta)
	\,{\rm d} r
\end{equation}
where $g(r, R, \beta) = 1 - 2 \beta R^2/r^2 + \beta(1+\beta) R^4/(2 r^4)$ and
\begin{equation}    \label{sol2}
	\nu \overline{v_r^4} (\beta={\rm const}) = 3 r^{-2 \beta}
	\int_r^\infty r^{2 \beta} \nu \sigma_r^2 (r)
	\frac{{\rm d} \Phi}{{\rm d} r} \ {\rm d} r \ .
\end{equation}
In the following we will rescale the fourth moment to obtain the line-of-sight or
projected kurtosis
\begin{equation}	\label{kurt}
	\kappa_{\rm los} (R) = \frac{\overline{v_{\rm los}^4} (R)}
	{\sigma_{\rm los}^4 (R)}
\end{equation}
whose value is 3 for a Gaussian distribution.

Having measured the line-of-sight velocities and projected positions for a number
of tracer particles or galaxies one can estimate the profiles of velocity dispersion
and kurtosis by taking $n$ tracer particles per bin and using the following
estimators of variance and kurtosis
\begin{equation}    \label{app1}
	S^2 = \frac{1}{n} \sum_{i=1}^n (v_i - \overline{v})^2
\end{equation}
and
\begin{equation}    \label{app2}
	K = \frac{\frac{1}{n} \sum_{i=1}^n (v_i - \overline{v})^4}{(S^2)^2}
\end{equation}
where
\begin{equation}    \label{app3}
	\overline{v} = \frac{1}{n} \sum_{i=1}^n v_i
\end{equation}
is the mean of velocities in a bin.
Since in the case of study of galaxy kinematics in clusters the number of galaxies usually
does not exceed a few hundred, and for our least sampled cluster in Table~\ref{properties}
(A0262) is as low as 120, in order to have at least 4 data points in each of the velocity
dispersion or kurtosis profiles we need to adopt a rather low number of $n=30$ objects
per bin. Using the Monte Carlo
method described in the Appendix of {\L}okas \& Mamon (2003) one
can then construct unbiased and Gaussian-distributed estimators of line-of-sight
velocity dispersion $s$ and kurtosis-like variable $k$
\begin{equation}    \label{app4}
	s = \left( \frac{n}{n-1} S^2 \right)^{1/2}
\end{equation}
\begin{equation}    \label{app5}
	k = \left[ \log \left( \frac{3}{2.68} K \right) \right]^{1/10}
\end{equation}
where $S$ and $K$ are given by equations (\ref{app1})-(\ref{app3}).
The factor $n-1$ in equation (\ref{app4}) is the well known correction for bias when
estimating the sample variance, valid independently of the underlying distribution.
In (\ref{app5}) the factor $3/2.68$ corrects for the bias in the kurtosis estimate,
i.e. unbiased estimate of kurtosis is $K' = 3K/2.68$, while
the rather complicated function of $K'$ assures that the sampling distribution of $k$ is
approximately Gaussian. We have also checked that defined in this way the measured
velocity dispersion and kurtosis in a given bin
are very weakly correlated so the data points can be
fitted as independent. The standard errors in the case of $s$ are
$s/\sqrt{2(n - 1)} $
while in the case of $k$ are approximately 0.02 (for $n \approx 30$).
In the following we
assign these sampling errors to our mock and real data points.

We will further assume that the total density distribution in the studied objects
(simulated haloes or galaxy clusters) and the distribution of the tracer are well
approximated up to the virial radius $r_v$
by the NFW formula (Navarro, Frenk \& White 1997)
\begin{equation}    \label{ro}
      \frac{\varrho(s)}{\varrho_{c,0}} = \frac{\Delta_c \,c^2 g(c)}{3 \,s\,(1+ c
     s)^2} \ ,
\end{equation}
where $s=r/r_v$, $\varrho_{c,0}$ is the present critical density,
$\Delta_c=101.9$ is the characteristic density parameter,
$c$ is the concentration parameter and $g(c) = [\ln (1+c) -
c/(1+c)]^{-1}$. We define the virial mass and radius as those with mean density
$\Delta_c=101.9$ times the critical density according to the spherical collapse
model for the standard $\Lambda$CDM cosmology (see {\L}okas \& Hoffman 2001). The
surface distribution of the tracer $I(R)$ following from 3D density profile
(\ref{ro}) can be found, together with other properties of the NFW haloes, in
{\L}okas \& Mamon (2001).

Having estimated the velocity dispersion and kurtosis profiles from the measured
positions and velocities we can fit these data with the solutions
(\ref{proj1})-(\ref{proj2}) estimating three free parameters: the virial mass $M_v$,
concentration $c$ and anisotropy $\beta$. In the remaining part of this section we
apply the method to a set of mock data in order to assess its viability in reproducing
the properties of real galaxy clusters. Our approach here is similar to that of
Sanchis et al. (2004), but in addition we address the problem of interlopers and
we estimate the errors in the parameters due to sampling errors.

\subsection{Removal of interlopers}

For this work we used the results of a cosmological dark matter simulation described by
Wojtak et al. (2005). The simulation was performed
within a box of size 150 $h^{-1}$ Mpc assuming
the concordance cosmological model ($\Lambda$CDM) with
parameters $\Omega_M=0.3$, $\Omega_{\Lambda}=0.7$, $h=0.7$ and $\sigma_8=0.9$.
We focused on 10 massive and isolated haloes extracted from the final output
of the simulation
whose properties are listed in Table~1 of Wojtak et al. (2005). In order to
emulate the observations we place an observer at a distance of
100 Mpc from a centre of a given halo so that he will be able to see it
receding with a velocity of around
7000 km s$^{-1}$. We assume that the line of sight is parallel to the x, y or z axis of
the simulation box, with respect to which the haloes should be oriented randomly. The
observer is located far enough from the halo so that the cone of observation can be
approximated by a cylinder. We then project all particle
velocities along the line of sight and
the distances on the surface of the sky restricting the observations to the circle of
projected radius $R=r_v$ on the sky. Next we reject all particles with velocities
differing from the mean velocity of the halo by more than $\pm 4000$ km s$^{-1}$,
as we did for
the real clusters. This cut-off corresponds to at least $4 \sigma_{\rm los}$ (with
$\sigma_{\rm los}$ calculated for all particles in the halo) so it is not very restrictive.
From the obtained sample of particles we randomly draw 300 out of about $10^4$ per halo.

\begin{figure}
\begin{center}
    \leavevmode
    \epsfxsize=8.3cm
    \epsfbox[100 30 480 580]{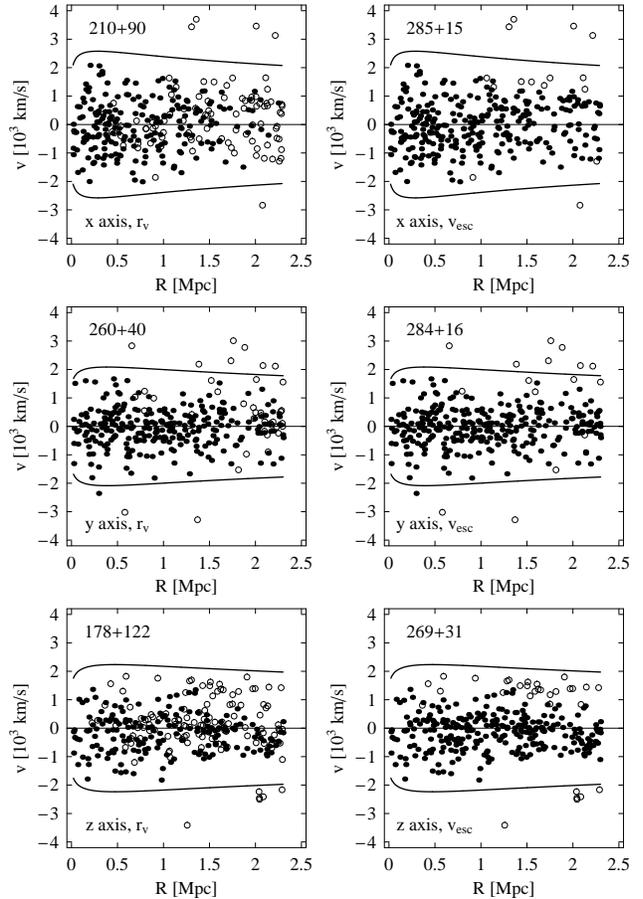}
\end{center}
\caption{Line-of-sight velocities of 300 particles sampled from halo 1
as a function of projected distance from the centre of the halo.
The upper, middle and lower row show observations performed
along $x$, $y$ and $z$ axis of the simulation box respectively. Filled (empty)
circles mark particles which actually reside inside (outside) $r_v$ (left column) or which
are bound (unbound) to the halo (right column). The numbers in the upper left corner
of each panel give the numbers of filled$+$empty circles.
Solid lines show fitted $\pm 3 \sigma_{\rm los} (R)$ profiles separating
the particles included in the analysis from those rejected as interlopers.}
\label{vrhalo1}
\end{figure}

The velocities and positions of 300 particles obtained in this way for halo 1
(of mass $M_v=7.5 \times 10^{14} M_{\sun}$ and virial radius $r_v=2.3$ Mpc) observed
along x, y and z axis (upper, middle and lower row respectively)
are shown in Fig.~\ref{vrhalo1}. In the plots shown in the left
column of the Figure the particles that happen to lie inside
the virial radius of the halo (as verified using 3D information)
were coded with filled circles, while those outside the virial radius are shown with
open symbols. In the right panel the coding is similar but the particles were
divided into those which are bound (have velocities smaller that the escape velocity,
$v < v_{\rm esc}$) or unbound to the halo. The axes along which the particles
were observed are marked in the lower left corner of each panel together with the labels
showing whether the distinction between the particles was made with the criterion of
$r_v$ or $v_{\rm esc}$. In the upper left corners of the plots we give the numbers of particles
fulfilling or not the criterion which sum to the total number of 300.

Averaging over 10 haloes and 3 directions of observation we find that
of the total number of particles (300), 76 percent reside inside $r_v$, 87 percent
inside 2 $r_v$, while 92 percent are actually bound to its halo. In addition, the unbound
particles are always a subsample of those with $r > 2 r_v$, i.e. there are no unbound
particles inside $2 r_v$.
The particles that do not fulfill any of the criteria are more common at larger
projected distances
from the centre of the halo. The particles from outside $r_v$ are obviously
candidates for interlopers since they would not be used to estimate the density profile
from the 3D information. However, as the cited numbers show, about half of them are
actually close to the halo (within 2 $r_v$) and most of them are bound to the halo
and therefore probably reasonably good tracers of the potential. This agrees with
the recent studies based on $N$-body simulations
(e.g. Klypin et al. 2003; Prada et al. 2005; Betancort-Rijo et al. 2005) which
demonstrate that the virialized region extends somewhat beyond the virial radius, as we
define it. Besides, the simulated dark matter haloes are usually not spherical and by
imposing spherical symmetry in our definition of virialized region we may in fact cut out
particles which are actually members of the halo. Anyway, most of the candidate interlopers
reside close to the mean velocity of the halo and only a few of them are true outliers
which could significantly alter our estimates of the velocity moments.

We proceed to remove these outliers in the following way. First we calculate the velocity
dispersion profiles by binning the data (with 30 particles per bin)
and assigning them sampling errors, as described
in the previous subsection. We then fit the data with solutions (\ref{proj1})
assuming $\beta=0$ and adjusting $M_v$ and $c$. Although the $\beta$ values of dark
matter haloes are mildly radial (with mean $\beta \approx 0.3$, see e.g. Wojtak et al.
2005; Mamon \& {\L}okas 2005) the actual values of the parameters are not very
important at this stage since
our purpose now is only to reproduce the shape of the dispersion profile and this can be
done well with 2 parameters instead of 3. If the velocity dispersion increases strongly
and the fit goes to values of $c<1$ (the NFW profile would not make sense since the scale
radius would be larger than the virial radius) we keep $c=1$ and adjust $M_v$ and $\beta$
instead. We then reject all particles lying outside the mean velocity
$\pm 3 \sigma_{\rm los} (R)$ where $\sigma_{\rm los} (R)$ is the velocity dispersion
profile obtained with our best-fitting parameters. The procedure is repeated until no
more particles are removed. In each iteration we also calculate new estimate of the mean
line-of-sight velocity of the particles with respect to which the rejection is performed.

The $\pm 3 \sigma_{\rm los} (R)$ profiles from the last
iterations are shown as solid lines in Fig.~\ref{vrhalo1}. As we can see, the procedure
removes the most obvious interlopers which affect the velocity moments most strongly. The mean
number of particles removed in all 30 experiments is 14 which corresponds to 61 percent of
the mean number of unbound particles per halo.
In rare cases where the initial velocity dispersion profile from the
data is strongly increasing the first fit may not remove any particles and it is
necessary to repeat the fitting for a smaller number of data points. In such cases it may also
happen that a member particle from the centre of the halo is removed. The effectiveness of the
method will be compared to other methods of interloper removal used in the literature
in a forthcoming paper (Wojtak et al., in preparation).

\subsection{Fitting of velocity moments}

\begin{figure}
\begin{center}
    \leavevmode
    \epsfxsize=7.2cm
    \epsfbox[95 25 290 590]{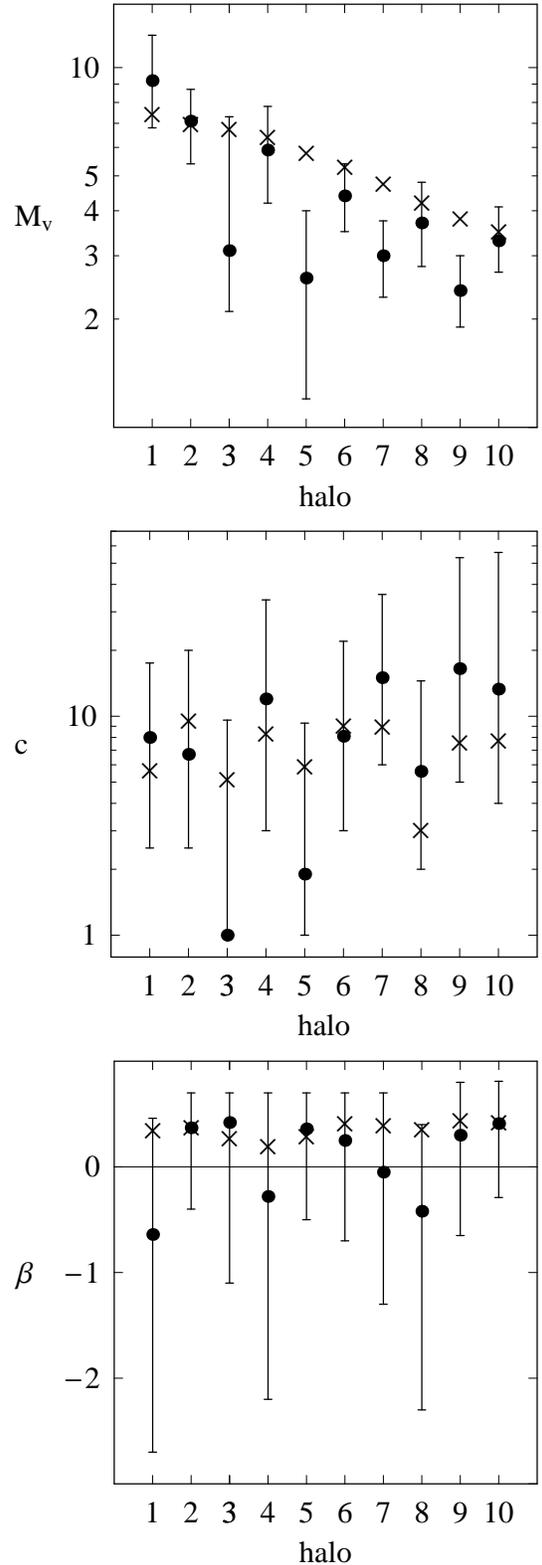}
\end{center}
\caption{Fitted parameters of the simulated dark matter haloes observed along the $x$ axis of
the simulation box. The three panels show
from top to bottom the virial mass $M_v$ (in units of $10^{14} M_{\sun}$), concentration
$c$ and the anisotropy parameter $\beta$. The filled dots show the parameters estimated from
the joint fitting of velocity dispersion and kurtosis data with $1\sigma$ error bars.
Crosses mark values estimated from the full 3D information.}
\label{fittedh}
\end{figure}

We now assess the viability of our method by jointly fitting both velocity dispersion
and kurtosis profiles calculated from our mock data for the
simulated dark matter haloes. After removal of interlopers we are left with 8-9 data
points for both dispersion and kurtosis which is of the same order as what we have for
real galaxy clusters. The quality of the obtained fits and magnitude of errors in the
parameters should therefore also be similar.

We jointly fit the mock data for velocity dispersion and kurtosis
with the predictions from the Jeans formulae (\ref{proj1})-(\ref{kurt}) adjusting the
three parameters $M_v$, $c$ and $\beta$. The parameters which minimize $\chi^2$ are shown
in Fig.~\ref{fittedh} as filled dots for 10 haloes observed along x axis of the simulation
box (for the observations along the y and z axis the results are similar).
The $1\sigma$ errors in the parameters due to the sampling errors, were found by
exploring the 3-dimensional
confidence region in the $M_v-\beta$, $\beta-c$ and $M_v-c$ parameter planes
and finding probability contours corresponding
to $1\sigma$, $2\sigma$ and $3\sigma$ i.e.
$\Delta \chi^2 = \chi^2 - \chi^2_{\rm min} = 3.53, 8.02, 14.2$.
The parameters of the haloes obtained from the full 3D information are marked as crosses.
The masses and concentrations were taken from Table~1 of Wojtak et al. (2005) and
the $\beta$ parameters were calculated from the $\beta$ profiles of the haloes by taking
averages in 10 radial bins inside the virial radius.

The discrepancies between the fitted and true values of the parameters may be due to
non-sphericity, presence of substructure, projected outliers and departures from
equilibrium e.g. in the form of streaming motions.
However, as we can see, the true values of the parameters are almost
always within the estimated $1\sigma$ error bars of the fitted values
(except for the mass estimates for haloes 5, 7 and 9 where the fitted values are
somewhat lower). We conclude from this analysis that the sampling errors are the main
source of error in this method.

We also note that the method might possibly work
even better for real galaxies in clusters than for our randomly selected particles.
Although it would be worthwhile to test the method on simulated galaxies, at the present
stage of the simulations this would be reduced to using subhaloes detected with standard
halo-finding techniques. The distributions of subhaloes both in space and velocity
are known to be biased with respect to those of dark matter particles (Diemand et al.
2004) and probably still suffer from overmerging problem. The density distribution
of subhaloes is flat in the centre while both the particles in
simulated haloes and galaxies in clusters have cuspy profiles. It would be very difficult
to disentangle the effects mentioned above from the uncertainties due to the use of
subhaloes.

Besides, simulations including baryons suggest that clusters tend to be
more spherical than pure dark matter haloes (Kazantzidis et al. 2004; Basilakos et al. 2005)
which would reduce the projection effects due to
non-sphericity of the systems. We also believe that cluster-cluster
mergers can be more easily detected
in real clusters than in a single final output of an $N$-body simulation of
dark matter haloes by studies of the X-ray emitting gas and
so our sample of clusters is probably more relaxed than the sample of
dark matter haloes we studied.
In the application of the method to real galaxy clusters we will therefore
neglect other sources of errors and estimate the uncertainties in the parameters
only from the sampling errors of the velocity moments.

\section{Results}

\subsection{Removal of interlopers}

\begin{figure}
\begin{center}
    \leavevmode
    \epsfxsize=8.3cm
    \epsfbox[100 40 480 580]{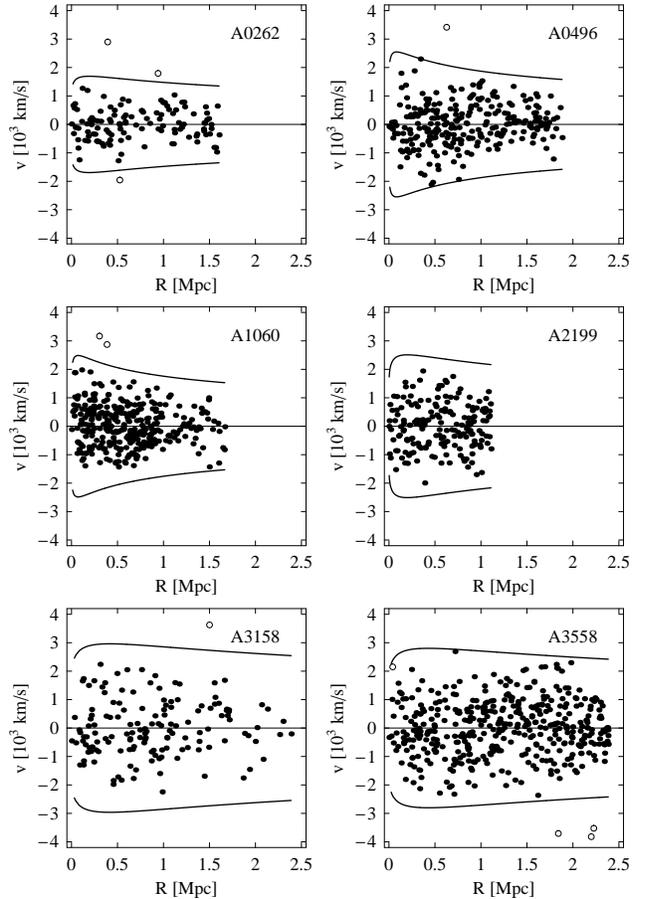}
\end{center}
\caption{Line-of-sight velocities of galaxies with respect to the cluster mean velocity
as a function of projected distance from the
cluster centre for the six clusters listed in Table~1.
Solid lines show fitted $\pm 3 \sigma_{\rm los} (R)$ profiles separating
the galaxies included in the analysis (filled circles)
from those rejected as interlopers (empty circles).}
\label{vr}
\end{figure}

We now proceed to apply the method of joint fitting of velocity moments to the sample
of six clusters listed in Table~\ref{properties}. We start by plotting in Fig.~\ref{vr}
the line-of-sight velocities of galaxies with respect to the cluster mean velocity
as a function of projected distance from the cluster centre. In order to separate the
galaxies which will be used in the calculation of velocity moments from the
supposed interlopers we apply the procedure for interloper removal described in the
previous section, exactly as we did for simulated data
with the additional assumption that the tracer density is proportional to
the total mass density (we do not infer the tracer density from the
observed surface number density of galaxies, because the latter would be very
uncertain for such small samples and probably
suffers from incompleteness that varies with projected radius).

Since we
do not a priori know the virial radius of the cluster, after estimating the virial mass
in each iteration we check whether all fitted data points lie inside the virial radius or
an additional data point could be included and adjust the number of points accordingly.
The $\pm 3 \sigma_{\rm los} (R)$ profiles obtained in the final iteration
of the procedure are
shown as solid lines in Fig.~\ref{vr}. As can be read from the Figure the number of
rejected galaxies is between 0 (for A2199) and 4 (for A3558), much lower than the number
of particles rejected during the application of the method to simulated haloes.
The rejected galaxy in the centre of A3558 lies inside the final $\pm 3 \sigma_{\rm los} (R)$
lines but was removed in the earlier iteration of the procedure.

\begin{figure}
\begin{center}
    \leavevmode
    \epsfxsize=8.3cm
    \epsfbox[100 40 480 580]{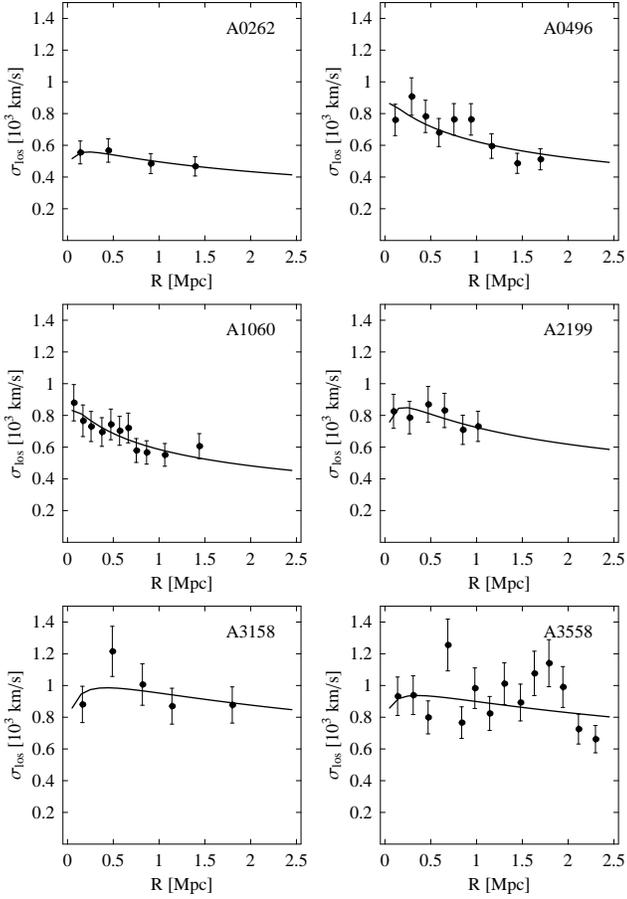}
\end{center}
\caption{Line-of-sight velocity dispersion as a function of projected distance from the
cluster centre for the six clusters. Solid lines show the best-fitting profiles
with parameters listed in Table~\ref{fittedpar}.}
\label{dispersion}
\end{figure}

\begin{figure}
\begin{center}
    \leavevmode
    \epsfxsize=8.3cm
    \epsfbox[100 40 480 580]{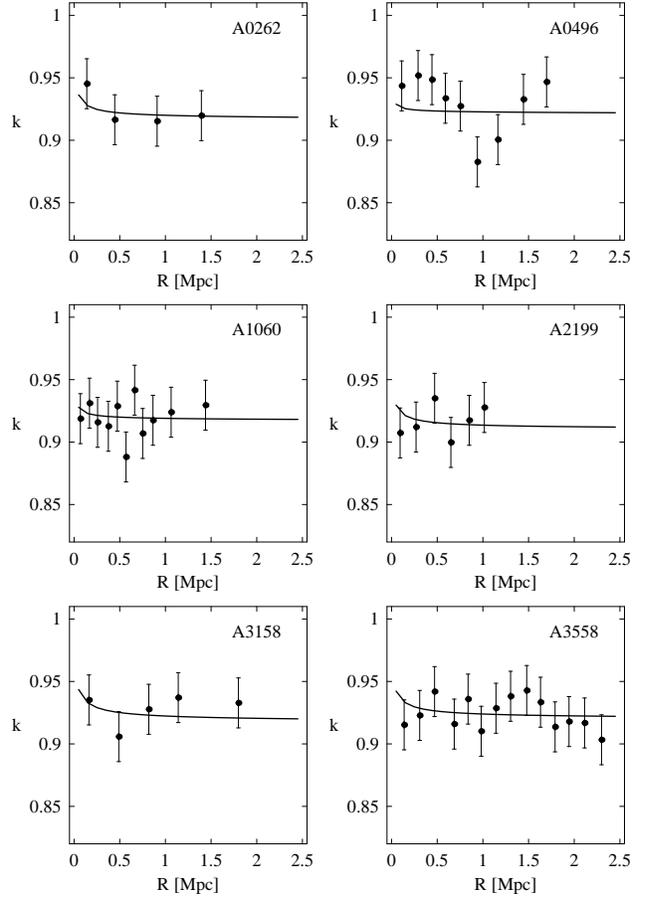}
\end{center}
\caption{Line-of-sight kurtosis variables as function of projected distance from the
cluster centre for the six clusters. Solid lines show the best-fitting profiles
with parameters listed in Table~\ref{fittedpar}.}
\label{kurtosis}
\end{figure}

Having found the final sample of galaxies we divide the data into radial bins of 30 galaxies
(except for A3158 which has 29 galaxies per bin) and calculate the velocity dispersion and kurtosis
profiles which are shown in Fig.~\ref{dispersion} and \ref{kurtosis}
with $1\sigma$ sampling errors. We see that the profiles are similar to those characteristic
for objects with NFW-like density profiles and orbits close to isotropic: the global trend is that
both profiles slightly decrease with radius (see {\L}okas \& Mamon 2001; Sanchis et al. 2004).
A3558 shows the most variable velocity dispersion profile with one discrepant point at 0.7 Mpc
(which is due to the single galaxy with discrepant velocity present in this bin --
see Fig.~\ref{vr}) and a secondary increase at about 1.8 Mpc while its kurtosis profile remains
rather uniform. On the other hand A0496, believed to be a very relaxed cluster,
has the most variable kurtosis profile. In an attempt to verify whether this variability may be
due to departures from equilibrium e.g. in the form of the presence of infalling groups of
galaxies we have also looked at the mean line-of-sight velocity profiles with respect to the
cluster mean. They do not however depart strongly from zero, typically remaining in each bin
within $0.2 V_v$
(the circular velocity at the virial radius as determined in the next subsection).

\subsection{Estimated parameters of the clusters}

\begin{figure}
\begin{center}
    \leavevmode
    \epsfxsize=7.2cm
    \epsfbox[95 25 290 590]{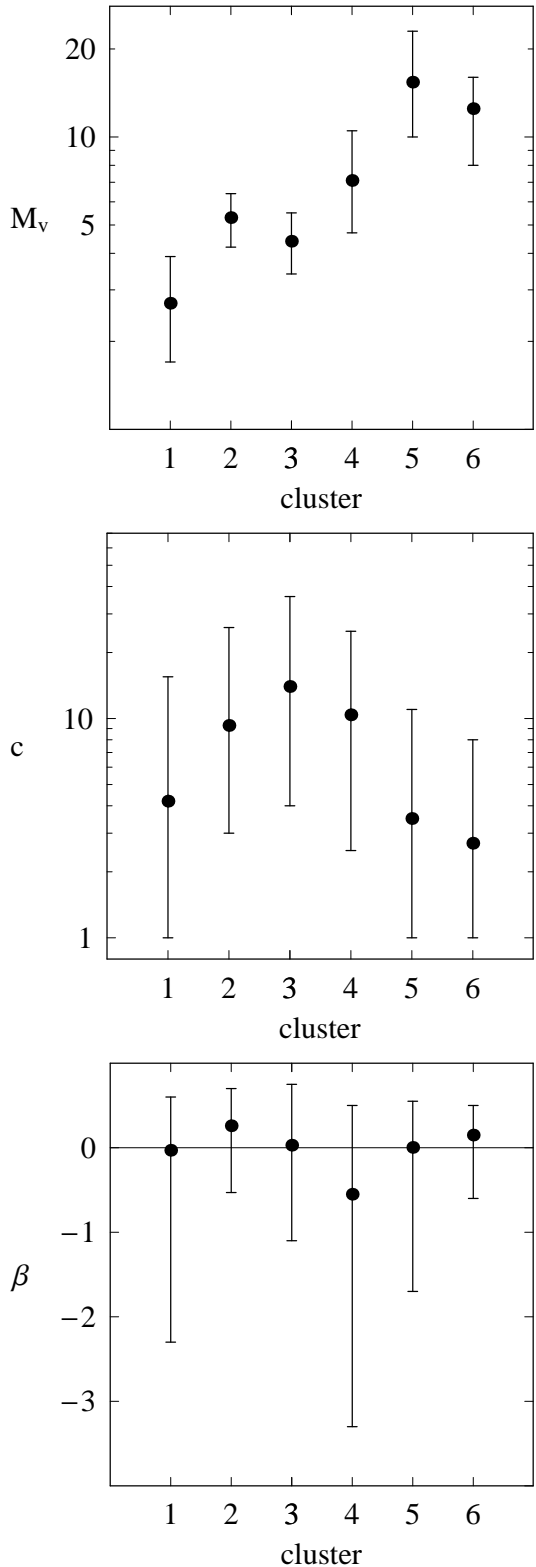}
\end{center}
\caption{Fitted parameters of the clusters. The three panels show
from top to bottom the virial mass $M_v$ (in units of $10^{14} M_{\sun}$), concentration
$c$ and the anisotropy parameter $\beta$ with $1\sigma$ error bars. The numbering of
clusters is the same as in Table~\ref{properties} and \ref{fittedpar}.}
\label{fitted}
\end{figure}

\begin{table}
\begin{center}
\caption{Fitted parameters of the clusters. }
\begin{tabular}{cccccc}
no.	 & cluster & $M_v $               & $c$ & $\beta$ & $\chi^2/N$ \\
	 &         & $[10^{14} M_{\sun}]$ &     &         &             \\
\hline
1 & A0262 & $2.7^{+1.2}_{-1.0}$ & $4.2^{+11.3}_{-3.2}$ & $-0.03^{+0.63}_{-2.27}$  & 1.1/5  \\    \\
2 & A0496 & $5.3^{+1.1}_{-1.1}$ & $9.3^{+16.7}_{-6.3}$ & $0.26^{+0.44}_{-0.79}$      & 18.5/15   \\   \\
3 & A1060 & $4.4^{+1.1}_{-1.0}$ & $14.0^{+22.0}_{-10.0}$ & $0.03^{+0.72}_{-1.13}$     & 8.6/19   \\   \\
4 & A2199 & $7.1^{+3.4}_{-2.4}$ & $10.4^{+14.6}_{-7.9}$ & $-0.55^{+1.05}_{-2.75}$      & 3.8/9   \\     \\
5 & A3158 & $15.4^{+7.6}_{-5.4}$ & $3.5^{+7.5}_{-2.5}$ & $0.004^{+0.55}_{-1.70}$    &  5.0/7     \\   \\
6 & A3558 & $12.5^{+3.5}_{-4.5}$ & $2.7^{+5.3}_{-1.7}$ & $0.15^{+0.35}_{-0.75}$     &  27.7/25    \\
\hline
\label{fittedpar}
\end{tabular}
\end{center}
\end{table}

In the following analysis we assume that the
galaxy distribution follows that of the total mass distribution. Although this assumption
is far from obvious there is evidence from studies of clusters (e.g. Carlberg et al. 1997;
{\L}okas \& Mamon 2003; Biviano \& Girardi 2003) that both number density and
luminosity density of galaxies in
clusters are cuspy and therefore can be quite well approximated by the NFW profile
(contrary to the distribution of subhaloes in the dark matter
simulations, which probably have a shallow core, see Diemand et al. 2004).
The concentration of these distributions does not have to be the same as that
of the total mass, but with numbers of galaxies as low as 120 per cluster we are not
able to reliably estimate the distribution of the tracer.

We jointly fitted the data for velocity dispersion and kurtosis
with the predictions from the Jeans formulae (\ref{proj1})-(\ref{kurt}) by minimizing
$\chi^2$ and adjusting the three parameters $M_v$, $c$ and $\beta$.
The best-fitting parameters for the six clusters are shown
in Fig.~\ref{fitted} as filled dots together with $1\sigma$ error bars.
The exact values are listed in Table~\ref{fittedpar} where the last column shows also
the goodness of fit measure, $\chi^2/N$. (The virial radii corresponding to the virial
masses are $r_v=1.2 [M_v/(10^{14} M_{\sun})]^{1/3}$ Mpc for our adopted cosmological
model.)
The errors in the parameters due to the sampling errors were found by
exploring the 3-dimensional
confidence region in the $M_v-\beta$, $\beta-c$ and $M_v-c$ parameter planes
and finding probability contours corresponding
to $1\sigma$, $2\sigma$ and $3\sigma$ i.e.
$\Delta \chi^2 = \chi^2 - \chi^2_{\rm min} = 3.53, 8.02, 14.2$.
The best-fitting velocity dispersion and kurtosis
profiles are shown in Fig.~\ref{dispersion} and \ref{kurtosis} as solid lines.

Interestingly, A3558 has the lowest concentration among all studied clusters. Although
the rather large error bars prevent us from concluding too much, this is expected since
A3558 is in the Shapley supercluster and therefore probably in the early stage of
cluster-cluster mergers which destroy the inner cusp. It is also worth noting that the
highest value of anisotropy ($\beta=0.26$) is obtained for A0496 which has the highest
value of line-of-sight kurtosis (see Table~1) as expected for non-rotating systems
(Merritt 1987).

\subsection{The mass-concentration relation}

\begin{figure}
\begin{center}
    \leavevmode
    \epsfxsize=7.2cm
    \epsfbox[90 10 290 210]{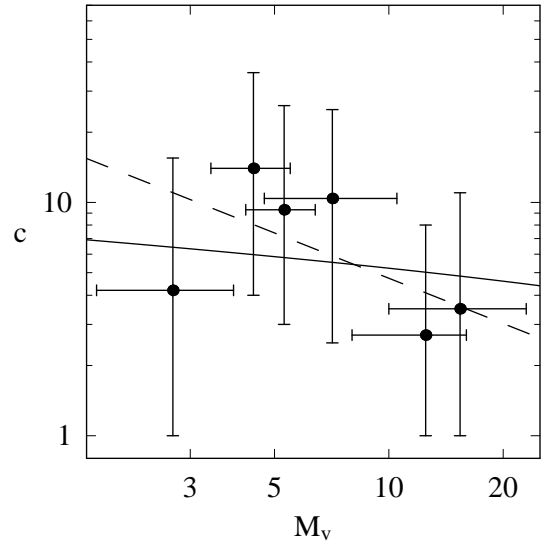}
\end{center}
\caption{The mass-concentration relation. The filled circles show the best-fitting
$M_v$-$c$ pairs for the six clusters with $1\sigma$ error bars. $M_v$ is given in
units of $10^{14} M_{\sun}$. The solid line
plots the prediction from the $N$-body simulations by Bullock et al (2001) while
the dashed line shows the best fit to the data points.}
\label{cm}
\end{figure}

The relation between the virial mass and concentration of dark matter haloes is a well
established result of $N$-body simulations and has been studied by many authors (e.g.
NFW; Bullock et al. 2001; Dolag et al. 2004).
In this subsection we address a question whether our best-fitting parameters of the clusters
agree with the trend of concentration decreasing with mass as found
in $N$-body simulations. We will compare our results to those of Bullock et al. (2001)
because they used the same definition of the virial radius.

In Fig.~\ref{cm} we plot estimated concentration $c$ as a function of cluster
virial mass $M_v$ (in units of $10^{14} M_{\sun}$)
for our six clusters. The $1\sigma$ errors following from sampling errors
of velocity moments were assigned to the points.
The solid line shows the approximation of the $c(M_v)$ relation calculated from the
toy model proposed by Bullock et al. (2001) which reproduces well the properties of
a large sample of haloes found in their $N$-body simulations. The predictions were obtained
for the standard cosmological model as in our simulations and $z=0$ from their equations
(9)-(13) with parameters $F=0.001$ and $K=3.0$ as advertised for masses
$M > 10^{14} h^{-1} M_{\sun}$.

Although the errors in the estimated parameters are quite
large it is clear from Fig.~\ref{cm}, that the data for clusters are consistent with
the $c(M)$ relation found in $N$-body simulations. To make this statement more
quantitative we have fitted to the data (neglecting the errors in mass) a linear
relation of the form $\log c = a \log [M_v/(10^{14} M_{\sun})] + b $ and found the
best-fitting parameters $a=-0.6 \pm 1.3$ and $b=1.3 \pm 1.1$ (at 68 percent
confidence level). This best-fitting relation is shown in Fig.~\ref{cm} as a dashed line.
Therefore, although the best-fitting slope is negative, the data
are also consistent at 1$\sigma$ level with constant concentration or even
concentration increasing with mass.

\subsection{The composite cluster}

\begin{figure}
\begin{center}
    \leavevmode
    \epsfxsize=7.1cm
    \epsfbox[95 35 290 590]{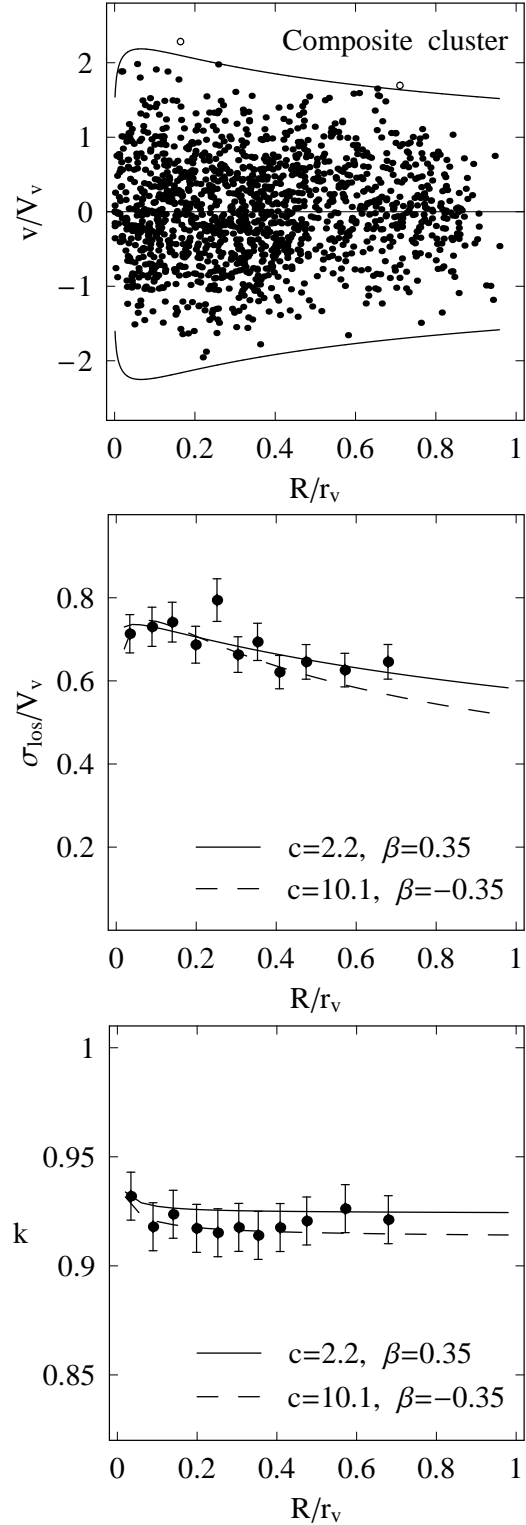}
\end{center}
\caption{Upper panel: line-of-sight velocities of 1465 galaxies of the composite cluster
normalized to $V_v$ as a function of projected distance in units of the corresponding virial radius.
Solid lines show fitted $\pm 3 \sigma_{\rm los} (R)$ profiles separating
the galaxies included in the analysis (filled circles)
from those rejected as interlopers (empty circles). Middle and lower panel:
line-of-sight velocity dispersion and kurtosis profiles for the composite cluster.
Solid and dashed lines show the best-fitting profiles with parameters listed in the
corners of the panels.}
\label{comp}
\end{figure}

\begin{figure}
\begin{center}
    \leavevmode
    \epsfxsize=7.2cm
    \epsfbox[95 20 290 390]{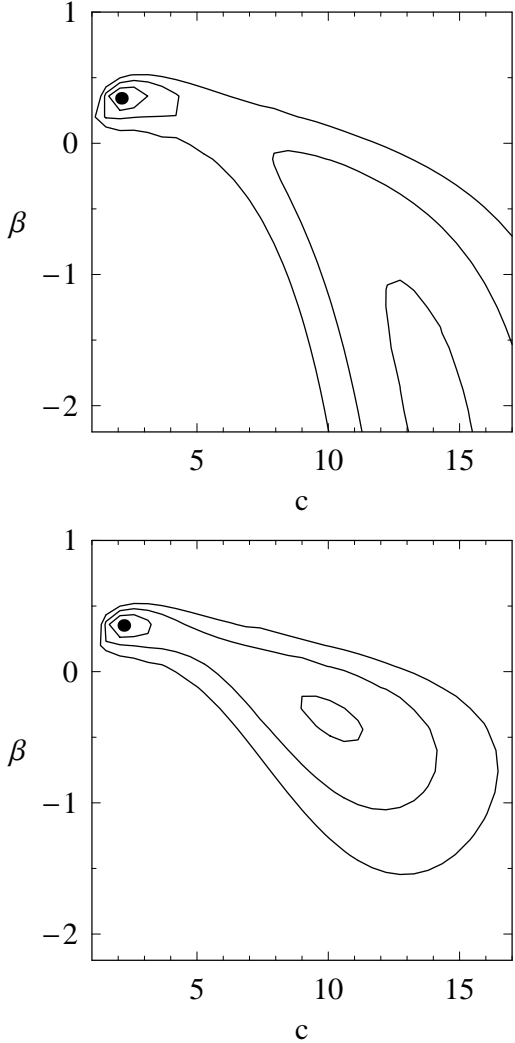}
\end{center}
\caption{The $1\sigma$, $2\sigma$ and $3\sigma$ probability contours in the $c-\beta$ parameter
plane obtained from fitting the velocity dispersion only (upper panel) and both velocity
dispersion and kurtosis (lower panel) of the composite cluster. The dots mark the best-fitting
parameters.}
\label{betac}
\end{figure}

The low number of galaxies per cluster in our sample results in rather large sampling errors
in the measured velocity moments and therefore also large errors in the estimated parameters
of the clusters. In order to reduce the uncertainties and study a typical cluster we have combined
the position and velocity data for our six objects to create a composite cluster.
We have normalized the galaxy distances by the estimated virial radius of their cluster, and
the velocities by the circular velocity at the virial radius $V_v=\sqrt{G M_v/r_v}$.
In this way we make the velocity moments independent of the virial mass.
The normalized velocities and velocity moments (calculated with 121 galaxies per
bin) for the composite cluster made of 1465 galaxies are shown
in Fig.~\ref{comp}. The kurtosis-like variable is now $k = [\log(3 K/2.90)]^{1/10}$
where the coefficient 2.90 was adjusted to the number of galaxies per bin.
The combined samples of galaxies were those after removal of interlopers in each cluster,
but we perform the procedure (fitting velocity dispersion profile and adjusting concentration
while keeping $\beta=0$) again on the total sample and it turns out that two more
interlopers have to be removed. The two galaxies have been marked as before with open circles
in the upper panel of Fig.~\ref{comp}. The two solid lines in this panel plot the last iteration
of $\pm 3 \sigma_{\rm los} (R)$ profiles separating the interlopers from the galaxies included
as members of the cluster.

With the mass dependence factored out, the velocity moments of the composite cluster depend
only on two parameters: concentration $c$ and anisotropy $\beta$. We adjust the parameters first
by fitting only the velocity dispersion profile. The best fit is then at $c=2.1$ and $\beta=0.34$ with
$\chi^2_{\rm min}/N= 6.4/9$. The $1\sigma$, $2\sigma$ and $3\sigma$ probability contours
corresponding
to $\Delta \chi^2 = \chi^2 - \chi^2_{\rm min} = 2.30, 6.17, 11.8$ are shown in the upper panel of
Fig.~\ref{betac} with the dot marking the pair of best-fitting parameters.
We can see that the $1\sigma$ confidence region is not restricted to the
vicinity of the best fit but appears also at more negative $\beta$ and larger $c$, towards
the second local minimum at $c=14.1$ and $\beta=-2.1$ with $\chi^2/N= 8.1/9$. Fitting both
the velocity dispersion and kurtosis breaks this degeneracy but only to some extent: the
best fit is now at $c=2.2$ and $\beta=0.35$ with $\chi^2_{\rm min}/N= 11.0/20$, but another local
minimum of $\chi^2/N=12.9/20$ is found at $c=10.1$ and $\beta=-0.35$.

The presence of the two
minima can be traced to the parameters estimated for each cluster separately in the last Section.
In our sample we had three clusters with concentration around 3 and three clusters with
concentrations of the order of 10. Those more concentrated also had lower mean value of $\beta$.
Indeed, performing similar fitting for composite clusters made separately of these two groups
we find single minima at $c=1.8$ and $\beta=0.2$ for low-concentration clusters while
$c=12.7$ and $\beta=-0.5$ for high-concentration clusters.

The profiles of the normalized moments for the best-fitting parameters are plotted as solid lines
in the middle and lower panel of Fig.~\ref{comp}. The solutions corresponding to the second
local $\chi^2$ minimum are plotted as dashed lines. Recall, that these are best-fitting profiles
obtained from the joint fit of dispersion and kurtosis and not to each of them separately --
this is why the fitted kurtosis profile does not match the corresponding data perfectly. Actually the
best fit to the kurtosis data alone would have $\beta$ closer to zero (isotropic orbits), but the
velocity dispersion forces $\beta$ towards more radial or more tangential values.
Therefore, to summarize our
results for the composite cluster, we find $1.5 < c < 14$ and $-1.1 < \beta < 0.5$ at the 95 percent
confidence level.

\section{Discussion}

We performed a joint analysis of velocity dispersion and kurtosis profiles of galaxies in six
nearby relaxed galaxy clusters estimating two global parameters of their mass distribution and
the anisotropy of galaxy orbits. The method of joint fitting the moments was tested on
simulated dark matter haloes and a new procedure for interloper removal was applied. With the
presently available number of galaxy redshifts per cluster the errors in the estimated parameters
are still large. The estimates of mass have an error from 20 percent in the case of A0496 to
50 percent in the case of A3158 and are in rough agreement with previous estimates (e.g. Girardi
et al. 1998; Rines et al. 2003).
The uncertainties in the estimates of concentration are even
larger. The fitted parameters are consistent with the mass-concentration relation
found in $N$-body simulations.

The method has been first applied a few years ago to the Coma cluster ({\L}okas \& Mamon 2003)
for which the largest number of galaxy redshifts is available. This larger amount of data allowed
for a more detailed analysis: the dark matter distribution could be modelled separately from
stars and gas, the luminosity density profile could be used as a tracer distribution without assuming
that galaxies trace the total or dark matter distribution, galaxies identified as members of binaries
could be removed from the sample and the analysis could be restricted to early type galaxies which
are believed to be more relaxed. For the present sample of galaxy clusters we had smaller numbers
of galaxies and had to simplify the analysis. However, we believe that the clusters discussed
here are really relaxed which appears not to be the case for Coma
(which is probably the product of a recent
merger as suggested by its perturbed gas distribution and the presence of two cD galaxies).
In addition, in the present study we have improved the method by introducing a reliable new procedure
for the removal of interlopers.

Inclusion of kurtosis in the analysis allowed us to constrain the anisotropy and thereby break
the degeneracy between the mass distribution and anisotropy present in the analysis of velocity
dispersion. This led us to expect that very tight constraints on the concentration parameter and
anisotropy could be obtained if the analysis was performed on a composite cluster made of
galaxies belonging to all six clusters. Although the anisotropy was indeed constrained also in
this case some degeneracy still persists between $c$ and $\beta$ and almost equally good fits
can be found for very different pairs of the two parameters. The degeneracy may however also be
due to the rather simple form of anisotropy which we assumed to be constant with radius. Although
isotropic orbits ($\beta=0$) can very well be a realistic case for virialized, mostly elliptical
galaxies in clusters, our samples contained also spirals which probably are on more radial orbits
(Biviano \& Katgert 2004). While isotropy might be preserved in the centre, our composite cluster
may be better described by an anisotropy parameter increasing with radius. This possibility will
be studied elsewhere.

The critical factor in studies of galaxy kinematics in clusters is the number of available
redshift measurements. Since the number of such measurements typically does not exceed a few
hundred, while the errors in the estimated parameters are mainly due to sampling errors, as
we have shown, the future of such studies will probably still be in analyzing composite
clusters. This does not allow however to study relations between individual objects, like the
mass-concentration or mass-temperature relation. Those have been reserved, till recently, to
studies based on X-ray observations. Although much tighter constraints e.g. on the concentration
parameter can be obtained in this case (Pointecouteau, Arnaud \& Pratt 2005),
these studies rely on assumptions like the hydrostatic equilibrium and rather
uncertain temperature profiles of the X-ray emitting gas. They should therefore be complemented by
extensive analysis of galaxy kinematics in clusters. This may soon become possible with systematic
photometric and spectroscopic surveys like WINGS (Fasano et al. 2005).

\section*{Acknowledgements}

We wish to thank M. Moles, K. Rines and the referee, A. Biviano, for their comments
on this work.
Computer simulations used in this paper were performed at the
Leibnizrechenzentrum (LRZ) in Munich.
E{\L} and RW are grateful for the hospitality of Astrophysikalisches
Institut Potsdam, Institut d'Astrophysique de Paris and
Instituto de Astrof{\'\i}sica de Andalucia where part of this work was done.
RW acknowledges the summer student program at Copernicus Center.
This research has made use of the NASA/IPAC Extragalactic Database (NED) which is
operated by the Jet Propulsion Laboratory, California Institute of Technology,
under contract with the National Aeronautics and Space Administration. We have also
used the LEDA database (http://leda.univ-lyon1.fr).
This work was partially supported by the
Polish Ministry of Scientific Research and Information Technology
under grant 1P03D02726 as well as
the Jumelage program Astronomie France Pologne of CNRS/PAN and the exchange program
of CSIC/PAN.


\begin{thebibliography}{}

\bibitem[]{bprzz} Bardelli S., Pisani A., Ramella M., Zucca E., Zamorani G., 1998,
	MNRAS, 300, 589
\bibitem[]{bpygt} Basilakos S., Plionis M., Yepes G., Gottl\"ober S., Turchaninov V.,
	2005, MNRAS, in press, astro-ph/0505620
\bibitem[]{bcpp} Betancort-Rijo J. E., Sanchez-Conde M. A., Prada F., Patiri S. G., 2005,
	submitted to ApJ, astro-ph/0509897
\bibitem[]{bm} Binney J., Mamon G. A., 1982, MNRAS, 200, 361
%\bibitem[]{bt} Binney J., Tremaine S., 1987, Galactic Dynamics. Princeton
%    Univ. Press, Princeton, chap. 4.
\bibitem[]{bg} Biviano A., Girardi M., 2003, ApJ, 585, 205
\bibitem[]{bk} Biviano A., Katgert P., 2004, A\&A, 424, 779
\bibitem[]{bs} Biviano A., Salucci P., 2005, submitted to A\&A, astro-ph/0511309
\bibitem[]{bu} Bullock J. S., Kolatt T. S., Sigad Y., Somerville R. S.,
    Kravtsov A. V., Klypin A. A., Primack J. R., Dekel A., 2001,
    MNRAS, 321, 559
\bibitem[]{carl} Carlberg R. G. et al. 1997, ApJL, 485, 13
\bibitem[]{dcrhm} Dantas C. C., de Carvalho R. R., Capelato H. V., Mazure A., 1997
	ApJ, 485, 447
\bibitem[]{dms} Diemand J., Moore B., Stadel J., 2004, MNRAS, 352, 535
\bibitem[]{dbp} Dolag K., Bartelmann M., Perrotta F., Baccigalupi C., Moscardini L.,
	Meneghetti M., Tormen G., 2004, A\&A, 416, 853
\bibitem[]{dagp} Durret F., Adami C., Gerbal D., Pislar V., 2000, A\&A, 356, 815
\bibitem[]{esca} Escalera E., Biviano A., Girardi M., Giuricin G., Mardirossian F.,
	Mazure A., Mezzetti M., 1994, ApJ, 423, 539
\bibitem[]{fasano} Fasano G., et al., 2005, A\&A, in press, astro-ph/0507247
\bibitem[]{fm} Fitchett M., Merritt D., 1988, ApJ, 335, 18
\bibitem[]{gir} Girardi M., Giuricin G., Mardirossian F., Mezzetti M., Boschin W.,
	1998, ApJ, 505, 74
\bibitem[]{goto} Goto T., 2005, MNRAS, 359, 1415
\bibitem[]{hor} Horner D. J., Baumgartner W. H., Gendreau K. C., Mushotzky R. F.,
	Loewenstein M., Molnar S. M., 2000, 197th AAS Meeting,
	Bulletin of the American Astronomical Society, 32, 1581
\bibitem[]{jf} Jones C., Forman W., 1999, ApJ, 511, 65
\bibitem[]{kbm} Katgert P., Biviano A., Mazure A., 2004, ApJ, 600, 657
%\bibitem[]{kmm} Kazantzidis S., Magorrian J., Moore B., 2004, ApJ, 601, 37
\bibitem[]{kkzanm} Kazantzidis S., Kravtsov A. V., Zentner A. R., Allgood B., Nagai D.,
	Moore B., 2004, ApJL, 611, 73
\bibitem[]{khkg} Klypin A., Hoffman Y., Kravtsov A. V., Gottl\"ober S., 2003,
	ApJ, 596, 19
\bibitem[]{lo} {\L}okas E. L., 2002, MNRAS, 333, 697
\bibitem[]{lh01} {\L}okas E. L., Hoffman Y., 2001, in
	Spooner N. J. C., Kudryavtsev V., eds, Proc. 3rd International Workshop, The
	Identification of Dark Matter. World Scientific, Singapore, p. 121
\bibitem[]{lm01} {\L}okas E. L., Mamon G. A., 2001, MNRAS, 321, 155
\bibitem[]{lm03} {\L}okas E. L., Mamon G. A., 2003, MNRAS, 343, 401
\bibitem[]{lmp} {\L}okas E. L., Mamon G. A., Prada F., 2005a, MNRAS, 363, 918
\bibitem[]{lpwmg} {\L}okas E. L., Prada F., Wojtak R., Moles M., Gottloeber S.,
	2005b, MNRAS Letters, in press, astro-ph/0507508
\bibitem[]{mg} Mahdavi A., Geller M. J., 2004, ApJ, 607, 202
\bibitem[]{ml} Mamon G. A., {\L}okas E. L., 2005, MNRAS, 363, 705
%\bibitem[]{mk} Merrifield M. R., Kent S. M., 1990, AJ, 99, 1548
\bibitem[]{me} Merritt D., 1987, ApJ, 313, 121
\bibitem[]{nfw} Navarro J. F., Frenk C. S., White S. D. M., 1997, ApJ, 490, 493
\bibitem[]{pap} Pointecouteau E., Arnaud M., Pratt G. W., 2005, A\&A, 435, 1
%\bibitem[]{prada} Prada F. et al., 2003, ApJ, 598, 260
\bibitem[]{pksbpgs} Prada F., Klypin A. A., Simonneau E., Betancort-Rijo J., Patiri S.,
	Gottloeber S., Sanchez-Conde M. A., 2005, submitted to ApJ, astro-ph/0506432
\bibitem[]{rines1} Rines K., Mahdavi A., Geller M. J., Diaferio A., Mohr J. J., Wegner G.,
	2001, ApJ, 555, 558
\bibitem[]{rines2} Rines K., Geller M. J., Diaferio A., Mahdavi A., Mohr J. J., Wegner G.,
	2002, AJ, 124, 1266
\bibitem[]{rines3} Rines K., Geller M. J., Kurtz M. J., Diaferio A., 2003, AJ, 126, 2152
\bibitem[]{slm} Sanchis T., \L okas E. L., Mamon G. A., 2004, MNRAS, 347, 1198
\bibitem[]{vdm} van der Marel R. P., Magorrian J., Carlberg R. G., Yee H.
    K. C., Ellingson E., 2000, AJ, 119, 2038
\bibitem[]{wlgm} Wojtak R., {\L}okas E. L., Gottl\"ober S., Mamon G. A., 2005,
	MNRAS, 361, L1


\end{thebibliography}
\end{document}